# Optical Stern-Gerlach effect beyond the rotating wave approximation

V E Lembessis

8 Aghiou Nikolaou Str., GR 18900 Salamis, Greece

E-mail : vlembessis@yahoo.com

**Abstract**

We show that the inclusion of counter-rotating terms, usually dropped, in the interaction Hamiltonian of the electric dipole of a two level atom with an electromagnetic field leads to significant modification of the splitting of an atomic beam known as Optical Stern Gerlach Effect which now acquires a "fine" structure.

**1. Introduction**

Interaction of laser light with atoms is the basis of atomic cooling and trapping and has paved the way for tremendous progress in various fields of atomic physics such as atom optics and Bose-Einstein condensation [1], [2]. The interaction depends strongly on the characteristics of the beam and the atom (which is considered as a two-level system) through two important parameters the Rabi frequency $\Omega$ and the detuning $\delta = \omega_0 - \omega_L$ (with $\omega_0$ the atomic transition frequency and $\omega_L$ the frequency of the beam). It is common place to consider an interaction Hamiltonian where the counter-rotating terms have been dropped. This is sufficient in case of resonant interaction or moderate detuning and Rabi frequencies.

Recent advances in optical trapping of neutral atoms are based on far-off resonance optical dipole traps [3]. In such case the counter rotating terms of the Hamiltonian become important and may have a substantial contribution to the trapping potential as has been shown both experimentally and theoretically [4], [5]. Moreover last years there is a series of studies of various problems in quantum optics beyond the RWA [6]-[11]. It is necessary to explore the effects of such terms in all mechanical effects of laser light on atoms where we may achieve large detuning or strong Rabi frequencies. In our work we are dealing with the deflection of an atom from a laser beam and specifically with the "revision" of the well known Optical Stern-Gerlach Effect [12], in the case where the full Hamiltonian is used.

In 1975 was pointed out that the trajectory of a two-level atom interacting with an optical field gradient can, under certain circumstances, be split into two paths, each path containing atoms in one of the two orthogonal states [12]. Because of the state selective deflection in field gradient, this effect is often refereed to as Optical Stern Gerlach effect (OSGE) and has been extensively studied and also experimentally demonstrated [12]-[15].

Deflection and splitting of an atomic beam has been experimentally observed by Moskowitz et al [14]. They have demonstrated momentum transfer from a near resonant standing wave field to a beam of sodium atoms. The atomic beam is split symmetrically into two peaks whose separation increases with field strength. The short interaction time ensures that spontaneous emission plays no role. In addition to splitting they have observed diffraction of the atomic beam due to exchange of even numbers of photons with the field. A more refined experiment has been done by Sleator et al [15]. In their work they realized the splitting of a metastable He atoms beam in the exact resonance case and the deflection in the non-resonant case. The difference between these experiments is that in the first case the atomic beam width was much larger than the standing-wave period and the atomic beam was split by the periodic potential into momentum states differing by multiples of $2\hbar k$. In the later case, the splitting arises from the exactly resonant interaction of the atomic beam with an optical field gradient, and the splitting angle is a continuous function of this field gradient. This is why the authors of this paper claim that they observe for a first time a genuine OSGE.

The theoretical analysis of OSGE has been based on the concept of dressed states. The dressed states are stationary states of the problem in the sense that they are not mixed by the oscillating field. They correspond to a physical picture of an atom surrounded ("dressed") by photons and interacting with them. The eigenvalues of these states can be viewed as the potential energies of the atom-plus-field considered as a single system and the resultant classical force is the gradient of this potential. The dressed atom picture was introduced by C. Cohen-Tannoudji [16], and is ideally suited model system for studying effects beyond RWA. It allows one to access both regimes in which RWA breaks down; namely large coupling and (locally) large detuning compared to resonance frequency. We refer particularly to a sequence of fluorescence processes which corresponds to a radiative cascade of the dressed atom downwards its energy diagram. Dressed states give rise to effects like the Autlet-Townes splitting [17], or electromagnetic induced transparency [18].

Careful discussion and extensive analysis of the OSGE has been done by Cook [13], and Kazantsev and colleagues [19]. Cook pointed out that the dressed interpretation is not always appropriate, in that the dressed states do not, in general, correspond to the split components of the atomic beam. It is only in the limit of $\sqrt{(\Omega^2 + \Delta^2)} \cdot t \gg 1$ ($\Omega$ being the Rabi frequency) where the correct results agree with the prediction of the dressed state formalism. Kazantsev and colleagues discuss in full detail the OSGE and emphasize the following: For non-zero detuning there is a critical value of detuning, which depends on the rate that the optical field turns on. If the detuning is larger than this critical value the atom is in either of the dressed states and not a superposition and sees a unique potential and is therefore simply deflected. For detuning smaller than the critical value because of non-adiabatic coupling the atom ends up in a superposition of dressed states and we observe splitting. So non-adiabaticity is the crucial point. This is ensured with an abrupt switching-on of the field. However we must point out that since non-RWA terms become important for very large detuning the condition of rapid switching-on of the field ($\tau_{swt}\Delta \ll 1$) is more difficult to realize.

## 2. Interaction of atom with the field

We study the optical dipole force in case of very large detuning or strong coupling. We follow a dressed state formalism for such a case. If we assume that the atom is in the ground state 1 and the field has $n$ photons then by absorbing a photon it will go up to the excited state 2 while the field is going to have $n$-1 photons and vice versa. In such a case we have the transitions shown by small arrows in the figure 1 between states $|1,n\rangle$ and $|2,n-1\rangle$. If however we do not consider the RWA then we must take into account the cases where the atom goes to the excited state by emitting a photon (see arrow connecting $|1,n\rangle$ with $|2,n+1\rangle$), or to the ground state by absorbing a photon (see arrow connecting $|2,n-1\rangle$ with $|1,n-2\rangle$).

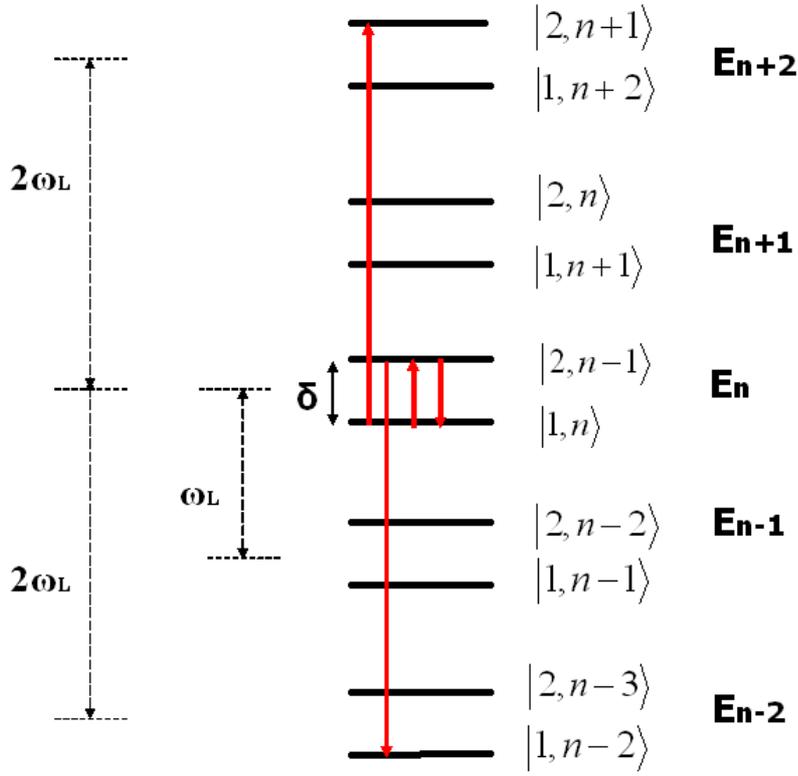

Figure 1. Atom-field energy manifolds. The larger red arrows correspond to the transitions induced by the non-RWA terms of the interaction Hamiltonian.

We assume that these four transitions now form a "closed family" of states accessible by the field and atom. This is an approximation, and so not quite true because if the system is found in manifolds $E_{n-2}$ or $E_{n+2}$ it may "jump" to other higher or lower manifolds as well. When off-resonant terms become significant each dressed state becomes a superposition of bare states from many manifolds. The justification of such an approximation will be shown later.

The Hamiltonian of this problem is given by:

$$H = \hbar\omega_0 |2,n-1\rangle\langle 2,n-1| + \hbar\omega_L a_L^+ a_L + i\hbar \{|2,n-1\rangle\langle 1,n|a_L - a_L^+|1,n\rangle\langle 2,n-1|$$

$$+ |1,n-2\rangle\langle 2,n-1|a_L - a_L^+|2,n+1\rangle\langle 1,n|\} \qquad (1)$$

Where $\omega_0$ is the atomic transition frequency and $\omega_L$ the frequency of the beam. The quantities $a_L$ and $a_L^+$ are the electric field destruction and creation operators respectively. Such a Hamiltonian may be represented by the following $4\times 4$ matrix.

$$\begin{bmatrix} n\hbar\omega_L & -i\hbar g_{\mathbf{k}}\sqrt{n} & -i\hbar g_{\mathbf{k}}\sqrt{n+1} & 0 \\ i\hbar g_{\mathbf{k}}\sqrt{n} & \hbar\omega_0 + (n-1)\hbar\omega_L & 0 & i\hbar g_{\mathbf{k}}\sqrt{n-1} \\ i\hbar g_{\mathbf{k}}\sqrt{n+1} & 0 & \hbar\omega_0 + (n+1)\hbar\omega_L & 0 \\ 0 & -i\hbar g_{\mathbf{k}}\sqrt{n-1} & 0 & (n-2)\hbar\omega_L \end{bmatrix} \quad (2)$$

Where we considered that $|1\rangle = |1,n\rangle, |2\rangle = |2,n-1\rangle, |3\rangle = |2,n+1\rangle$ and $|4\rangle = |1,n-2\rangle$. Note that the above Hamiltonian depends on the number $n$ of photon in the manifold but we shall neglect it by supposing that the laser beam is initially excited in a coherent state with Poisson distribution for $n$, the width $\Delta n$ of which is very small compared with the average number of photons [16]. We are not interested in the change of the field during the coupling and assume $n$ to be very large. If $\delta = \omega_0 - \omega_L$ is the detuning and assume that in our case the photon number $n$ is very large then the above matrix becomes:

$$\hbar \cdot \begin{bmatrix} n\omega_L & -ig_{\mathbf{k}}\sqrt{n} & -ig_{\mathbf{k}}\sqrt{n} & 0 \\ ig_{\mathbf{k}}\sqrt{n} & n\omega_L + \delta & 0 & ig_{\mathbf{k}}\sqrt{n} \\ ig_{\mathbf{k}}\sqrt{n} & 0 & n\omega_L - \delta + 2\omega_0 & 0 \\ 0 & -ig_{\mathbf{k}}\sqrt{n} & 0 & n\omega_L + 2\delta - 2\omega_0 \end{bmatrix} \quad (3)$$

Here for simplicity we consider the case where the electric field is a standing wave which corresponds to a real $g_{\mathbf{k}}$. The eigenvalues $\omega$ of this matrix will give us the energies of the 4 dressed states. To facilitate calculations I consider the following transformation $\omega = n\omega_L + x, \beta = g_{\mathbf{k}}\sqrt{n}$. The eigenvalue determinant takes the form:

$$\begin{bmatrix} -x & -i\beta & -i\beta & 0 \\ i\beta & \delta - x & 0 & i\beta \\ i\beta & 0 & 2\omega_0 - \delta - x & 0 \\ 0 & -i\beta & 0 & 2\delta - 2\omega_0 - x \end{bmatrix} = 0 \quad (4)$$

The solution of such eigenvalue problem can be found both analytically and numerically but there are some points which may facilitate our calculations. In real problems of FORT the detuning is much higher than the Rabi frequency [4]. In this case the determinant above becomes diagonal and the eigenvalues are easily found:

$$x_1 = 0, \ x_2 = \delta, \ x_3 = 2\omega_0 - \delta \text{ and } x_4 = 2\delta - 2\omega_0 \quad (5)$$

In such case it is easy to take into account more transitions to higher and lower manifolds and to calculate the splitting frequencies. In the generic case the field+atom system is found in states: $|1, n+k\rangle$ or $|2, n+k-1\rangle$ with $k = 0, \pm 2, \pm 4,...$. Note that $k$ is an even number; this is the "selection" rule for transitions in the "ladder" of the system's states once RWA and spontaneous emission are not taken into account. It is the spontaneous emission that would make all the states of the "ladder" accessible by the system. This means, that although the dressed states are superpositions of all involved bare states, a selection rule between undressed states exists. This differs from spontaneous decay in optical dressing, where transitions between all dressed states can occur. Creating the matrix of the interaction Hamiltonian we get its diagonal elements from the following state products

$$\langle 1, n+k | 1, n+k \rangle = n\hbar\omega_L - k\omega_L \tag{6}$$

$$\langle 2, n+k-1 | 2, n+k-1 \rangle = n\hbar\omega_L + \omega_0 + (k-1)\omega_L \tag{7}$$

Since the determinant of the interaction matrix is diagonal the above simple relations give us the whole spectrum of the splitting frequencies. For the non-diagonal elements of the interaction matrix which represent transitions among different levels we have

$$\langle 1, n+k | 2, n+k-1 \rangle = -i\hbar g \sqrt{n} \tag{8}$$

$$\langle 1, n+k | 2, n+k+1 \rangle = -i\hbar g \sqrt{n+1} \tag{9}$$

Let's now turn to the generic problem. The solution of such eigenvalue problem leads to the solution of a quartic equation. A full analytical solution of such an equation has been given by Ferrari [20] and leads to the following eigenvalues:

$$x_1 = \frac{\delta}{2} + \frac{1}{2}\left[6\beta^2 + a + 2\left(5\beta^4 + c\beta^2 + d\right)^{1/2}\right]^{1/2} \tag{10}$$

$$x_2 = \frac{\delta}{2} - \frac{1}{2}\left[6\beta^2 + a + 2\left(5\beta^4 + c\beta^2 + d\right)^{1/2}\right]^{1/2} \tag{11}$$

$$x_3 = \frac{\delta}{2} + \frac{1}{2}\left[6\beta^2 + a - 2\left(5\beta^4 + c\beta^2 + d\right)^{1/2}\right]^{1/2} \tag{12}$$

$$x_4 = \frac{\delta}{2} - \frac{1}{2}\left[6\beta^2 + a - 2\left(5\beta^4 + c\beta^2 + d\right)^{1/2}\right]^{1/2} \tag{13}$$

with the quantities $a, c, d$ given by

$$a = 5\delta^2 - 12\delta\omega_0 + 8\omega_0^2$$

$$c = 12\delta^2 + 8\omega_0^2 - 20\delta\omega_0$$

$$d = 52\delta^2\omega_0^2 - 48\delta\omega_0^3 - 24\delta^3\omega_0 + 16\omega_0^4 + 4\delta^4.$$

The above four eigenvalues correspond to the following normalized eigenstates (dressed states)

$$\psi_i = \left(a_{i1}^2 + |a_{i2}|^2 + |a_{i3}|^2 + a_{i4}^2\right)^{-1/2} \left\{a_{i1}|1,n\rangle + a_{i2}|2,n-1\rangle + a_{i3}|2,n+1\rangle + a_{i4}|1,n-2\rangle\right\} \tag{14}$$

where the coefficients inside the curly brackets are given by:

$$a_{i1} = -\left(\beta^2 - 2x_i\omega_0 + 3x_i\delta - 2\delta^2 + 2\delta\omega_0 - x_i^2\right)/\beta^2$$

$$a_{i2} = i(-2\delta + x_i)/\beta$$

$$a_{i3} = i\left(-2\beta^2\omega_0 + 2\delta\beta^2 - 2x_i\beta^2 - 2x_i\delta\omega_0 - 3\delta x_i^2 + 2x_i^2\omega_0 + 2x_i\delta^2 + x_i^3\right)/\beta^3$$

$$a_{i4} = 1$$

for $i$ = 1, 2, 3, 4

In the case of a position dependent Rabi frequency the above eigenfrequencies lead to the following forces on the atom $\mathbf{F}_i = -\hbar\vec{\nabla}x_i$ with $i$ = 1, 2, 3, 4,

$$\mathbf{F}_1 = 2\beta\vec{\nabla}\beta\left[6\beta^2 + a + 2\left(5\beta^4 + c\beta^2 + d\right)^{1/2}\right]^{-1/2}\left[6 + \left(10\beta^2 + c\right)\left(5\beta^4 + c\beta^2 + d\right)^{-1/2}\right] \quad (15)$$

$$\mathbf{F}_2 = -2\beta\vec{\nabla}\beta\left[6\beta^2 + a + 2\left(5\beta^4 + c\beta^2 + d\right)^{1/2}\right]^{-1/2}\left[6 + \left(10\beta^2 + c\right)\left(5\beta^4 + c\beta^2 + d\right)^{-1/2}\right]$$

(16)

$$\mathbf{F}_3 = 2\beta\vec{\nabla}\beta\left[6\beta^2 + a + 2\left(5\beta^4 + c\beta^2 + d\right)^{1/2}\right]^{-1/2}\left[6 - \left(10\beta^2 + c\right)\left(5\beta^4 + c\beta^2 + d\right)^{-1/2}\right]$$

(17)

$$\mathbf{F}_4 = -2\beta\vec{\nabla}\beta\left[6\beta^2 + a + 2\left(5\beta^4 + c\beta^2 + d\right)^{1/2}\right]^{-1/2}\left[6 - \left(10\beta^2 + c\right)\left(5\beta^4 + c\beta^2 + d\right)^{-1/2}\right]$$

(18)

These four forces lead to a splitting of the atomic wave packet in four parts instead of two which we get once RWA is made. With these new terms the "spectrum" of the wave-packet presents a "fine" structure. Such a structure has been predicted if the electric field has a statistical distribution [21]. In that case the authors, who consider RWA, take into account the quantum nature of the laser light in the optical Stern-Gerlach effect and predict a multipeaked form of the deflection profile (fine structure) which is attributed to the multilevel nature of the two-level atom dressed states and depends on the statistics of the field. In our case, though, the multipeaked nature of the splitting is due to the inclusion of non-RWA terms in the interaction. Experiments on OSGE were performed in the regime of exact resonance or moderate detunings and weak coupling so this "fine" structure was not observed [14], [15].

### 3. Illustrations

We are going to give now some numerical examples which show the modification of the OSGE. We consider the following cases concerning the excitation of the transition $5^2S_{1/2} - 5^2P_{3/2}$ in an $^{85}Rb$ atom where $\omega_o = 3.2\times10^{15}\,r/s$ and the excited state linewidth is $\Gamma = 37.6\times10^6$ $r/s$:

A) Let the case where $\Delta = 0$ and $\Omega = 3.76 \times 10^9 \, r/s$ (100Γ). In this case the effect of non-RWA terms is negligible. In this case we must consider an interaction time such that spontaneous emission plays no role. The four eigenvalues are $x_1 = 0.184 \times 10^{10} \, r/s$, $x_2 = -0.184 \times 10^{10} \, r/s$, $x_3 = 0.64 \times 10^{16} \, r/s$, $x_4 = -0.64 \times 10^{16} \, r/s$. The first two of them correspond to the ordinary eigenvalues given when we make the RWA in the interaction Hamiltonian. The four dressed states are given by:

$$\psi_1 = -0.707|1,n\rangle + 0.707i|2,n-1\rangle + 0.2 \times 10^{-6} i|2,n+1\rangle + 0.2 \times 10^{-6}|1,n-2\rangle$$

$$\psi_2 = -0.707|1,n\rangle - 0.707i|2,n-1\rangle + 0.2 \times 10^{-6} i|2,n+1\rangle - 0.2 \times 10^{-6}|1,n-2\rangle$$

$$\psi_3 = -0.29 \times 10^{-6}|1,n\rangle - 0.3 \times 10^{-9} i|2,n-1\rangle + i|2,n+1\rangle + 0.71 \times 10^{-16}|1,n-2\rangle$$

$$\psi_4 = 0.82 \times 10^{-13}|1,n\rangle - 0.29 \times 10^{-6} i|2,n-1\rangle + 0.69 \times 10^{-16} i|2,n+1\rangle + |1,n-2\rangle$$

B) Let the case where $\Delta = 0$ and $\Omega = 3.76 \times 10^{12} \, r/s$ ($10^5$ Γ). In this case the effect of non-RWA terms starts to be considerable. In this case we must also consider an interaction time such that spontaneous emission plays no role. The four eigenvalues are $x_1 = 0.183 \times 10^{13} \, r/s$, $x_2 = -0.184 \times 10^{13} \, r/s$, $x_3 = 0.64 \times 10^{16} \, r/s$, $x_4 = -0.64 \times 10^{16} \, r/s$. The corresponding dressed states are

$$\psi_1 = -0.707|1,n\rangle - 0.707i|2,n-1\rangle + 0.2 \times 10^{-3} i|2,n+1\rangle + 0.2 \times 10^{-3}|1,n-2\rangle$$

$$\psi_2 = 0.707|1,n\rangle - 0.707i|2,n-1\rangle - 0.2 \times 10^{-6} i|2,n+1\rangle - 0.2 \times 10^{-6}|1,n-2\rangle$$

$$\psi_3 = 0.29 \times 10^{-3}|1,n\rangle + 0.83 \times 10^{-7} i|2,n-1\rangle + 0.99i|2,n+1\rangle + 0.37 \times 10^{-11}|1,n-2\rangle$$

$$\psi_4 = 0.82 \times 10^{-7}|1,n\rangle - 0.29 \times 10^{-3} i|2,n-1\rangle - 0.11 \times 10^{-10} i|2,n+1\rangle + 0.99|1,n-2\rangle$$

C) Let the case where $\Delta = 0.1 \times 10^{12} \, r/s$ ($2.7 \times 10^4$ Γ) and $\Omega = 3.67 \times 10^{12} \, r/s$ ($10^5$ Γ). In this case the effect of non-RWA terms begin to be considerable. Detuning is quite larger than the spontaneous emission rate so spontaneous emission in any time scale has no effect. The four eigenvalues are $x_1 = 0.189 \times 10^{13} \, r/s$, $x_2 = -0.179 \times 10^{13} \, r/s$, $x_3 = 0.64 \times 10^{16} \, r/s$, $x_4 = -0.64 \times 10^{16} \, r/s$.

The corresponding dressed states are

$$\psi_1 = -0.697|1,n\rangle - 0.717i|2,n-1\rangle + 0.2 \times 10^{-3} i|2,n+1\rangle + 0.2 \times 10^{-3}|1,n-2\rangle$$

$$\psi_2 = -0.717|1,n\rangle + 0.697i|2,n-1\rangle + 0.2 \times 10^{-6} i|2,n+1\rangle + 0.2 \times 10^{-6}|1,n-2\rangle$$

$$\psi_3 = 0.29 \times 10^{-3}|1,n\rangle + 0.82 \times 10^{-7} i|2,n-1\rangle + 0.99i|2,n+1\rangle + 0.432 \times 10^{-11}|1,n-2\rangle$$

$$\psi_4 = 0.82 \times 10^{-7} |1,n\rangle - 0.29 \times 10^{-3} i |2, n-1\rangle - 0.11 \times 10^{-10} i |2, n+1\rangle + 0.99 |1, n-2\rangle$$

All the above illustrations show clearly that as detuning or Rabi frequency get larger then the splitting of the wavefunction acquires a "fine" structure due to the influence of the non-RWA terms of the total Hamiltonian. Knowing the dressed states, the time evolution of the system is straightforward [22]. Let's assume that the system is initially in the ground state with $n$ photons, i.e. $|\psi_I(0)\rangle = |1,n\rangle$. The initial state can be expressed in terms of dressed states. Each dressed state evolves in time according to

$$\psi_i(t) = \psi_i \exp(-i x_i t). \tag{19}$$

The evolution of the system, i.e. the state of the system after time $t$ is given by

$$|\psi_I(t)\rangle = \frac{D_1}{D} \tag{20}$$

with

$$D_i = \begin{vmatrix} \psi_1 e^{-x_1 t} & a_{12} & a_{13} & a_{14} \\ \psi_2 e^{-x_2 t} & a_{22} & a_{23} & a_{24} \\ \psi_3 e^{-x_3 t} & a_{32} & a_{33} & a_{34} \\ \psi_4 e^{-x_4 t} & a_{42} & a_{43} & a_{44} \end{vmatrix} \tag{21}$$

and

$$D_i = \begin{vmatrix} a_{11} & a_{12} & a_{13} & a_{14} \\ a_{21} & a_{22} & a_{23} & a_{24} \\ a_{31} & a_{32} & a_{33} & a_{34} \\ a_{41} & a_{42} & a_{43} & a_{44} \end{vmatrix} \tag{22}$$

**4. Conclusions**

We have shown that inclusion of the non-RWA terms in the interaction Hamiltonian between a two-level atom and a laser beam gives rise to a "fine" structure in the splitting of the atomic wave-function in the Optical Stern-Gerlach Effect. Our work was based on the dressed states formalism which is a proper choice for the case of large detuning or strong coupling. The effects of spontaneous emission have not been taken into consideration. As experimental work with Far Off-Resonance Traps for atomic particles grows, we consider that several mechanical effects of light should be revised by including the full Hamiltonian into the theoretical calculations and the corresponding experimental work.

**Acknowledgments**
This work was supported by the Research Networking Program "Quantum Degenerate Dilute Systems" (QUDEDIS) of the European Science Foundation. A large part of the work has been done while I was a host in the group of Pr. D. Meschede in Bonn, whom I am grateful to. I would like also to thank Pr. L. Allen for his valuable suggestion and remarks on the manuscript.